\documentclass[%
 reprint,
 amsmath,amssymb,
 aps,
]{revtex4-1}

\usepackage[pdftex]{graphicx}
\usepackage{dcolumn}
\usepackage{bm}
\usepackage{multirow}

\begin{document}

\title{Optimal Cooperative Searching Using Purely Repulsive Interactions}

\author{Noriyuki P. Tani}
 \author{Alan Blatt}
\author{David A. Quint}%
\email{dquint@ucmerced.edu}
\author{Ajay Gopinathan}
\email{agopinathan@ucmerced.edu}
\affiliation{Physics Department, University of California Merced}

\preprint{APS/123-QED}

\begin{abstract}
Foraging, either solitarily or collectively, is a necessary behavior for survival that is demonstrated by many organisms. Foraging can be collectively optimized by utilizing communication between the organisms. Examples of such communication range from high level strategic foraging by animal groups to rudimentary signaling among unicellular organisms. Here we systematically study the simplest form of communication via long range repulsive interactions between two diffusing Brownian searchers on a one-dimensional lattice. We show that the mean first passage time for either of them to reach a fixed target depends non-monotonically on the range of the interaction and can be optimized for a repulsive range that is comparable to the average spacing between searchers. Our results suggest that even the most rudimentary form of collective searching does in fact lower the search time for the foragers suggesting robust mechanisms for search optimization in cellular communities.
\end{abstract}

\pacs{Valid PACS appear here}
\maketitle
\section*{Introduction}
Understanding the process of searching or foraging in living systems has been of great interest in many disciplines, such as biology, physics, computer science, and robotics. The mechanisms by which different organisms forage for food can be quite varied, for example bears and wolves use their sense of smell in order to acquire food~\cite{Mattson}, while bats and dolphins use echolocation to locate their food~\cite{Au,Schnitzler}. Some animals have the ability to search for food individually; however, many other organisms must work in tandem in order to efficiently find food, such as ants and fish ~\cite{Ioannou,Jackson}. Studying collective foraging patterns in nature can reveal basic algorithmic features that can be directly compared with artificial searching algorithms used in computer science and robotics~\cite{Shoghian}. This type of analysis can help animal behavioral scientists and computer scientists understand how these algorithms evolved over time and became robust over the wide range of environmental scenarios.  One application of these searching process is currently used in robotics, where robots can utilize collective searching motifs that help them navigate unexplored terrain and also assist in search and rescue efforts ~\cite{Saeedi,Ko,Reich,Hoff}. Cooperation among both living systems and artificial ones strive for the same goals, such as minimizing the search time (i.e minimize energetic cost) while maximizing the search space.

Collective foragers or searchers, found in nature, display a high degree of coordination and communication within the collective as compared to a single searcher on its own. In fact, movement at the individual organism level within a collective is strongly correlated with the information that is being derived from their surrounding neighbors. A model system, which spans both types of foraging behavior, is the eukaryotic cell Dictyostelium discoideum (dicty). In a nutrient rich environment, single cell dicty move more or less randomly in the search for food, because in such an environment movement in any direction will yield an intake of food~\cite{Gole}. This type of behavior or type of locomotion may be regarded in all aspects as Brownian motion, where single foragers move randomly through their environment~\cite{Sheetz,Selmeczi}. In contrast,  when food supplies become scarce, collective foraging becomes more energy efficient at the single cell level~\cite{Gelbart}. In order to collectively forage, dicty cells communicate with each other through repulsive and attractive chemotactic signaling~\cite{Keating,Konijn,Pan}. When dicty displays this type of behavior we may consider them as interacting Brownian particles, where individual dicty motion is less random than in the nutrient rich environment and becomes more correlated with other dicty cells in the environment. This correlation is mainly due to the chemical signaling that is taking place, since dicty tries to avoid searching areas that were previously covered by avoiding the chemo-repellent areas left by other dicty cells, as well as moving toward chemo-attractive signals, where there may be a high local concentration of nutrients. Previous studies of short range interacting Brownian particles have been used to model biological systems such as transcription proteins sliding along DNA~\cite{Ryabov} and the swarming behaviors observed in birds and fishes \cite{Markis,Katz,Quint}. 

In this study, we are interested in how interacting Brownian foragers cooperate with each other, while searching for a single target (ex. food). This minimal model will allow us to distill out the necessary mechanisms and advantages of collective foraging. We address these questions by simulating two Brownian particles that search for a fixed target on a closed one dimensional lattice. In the simplest case we first study this system without any interaction between the two searchers and then compare this with the more complex system of two interacting Brownian particles by measuring the average Mean First Passage Time (MFPT) to the target~\cite{Cepa,Sokolov}. We found that interactions among the searchers affected the search time, and an optimal repulsion for foraging was found. This suggests that in order to optimize collective foraging, organism should interact such that they minimize redundant search patterns and maximize the search area in their environment. In section 1, we discuss our model and the dynamics of our simulation. In section 2 we present our results; results subsection I we compare the MFPT of three different systems; one searcher, two searchers without interactions, and two searchers with interactions; results subsection II, we present, by dimensional analysis, the relationship between the optimal repulsive strength and the lattice size; results subsection III, we present the relationship between the average encounter time and the lattice size. In section 3, we discuss the implications of our results. 

\section*{1. Model and Simulation}
\begin{figure}[ht]
\centering
\includegraphics[trim=0 0 0 1, clip=true, scale=0.25]{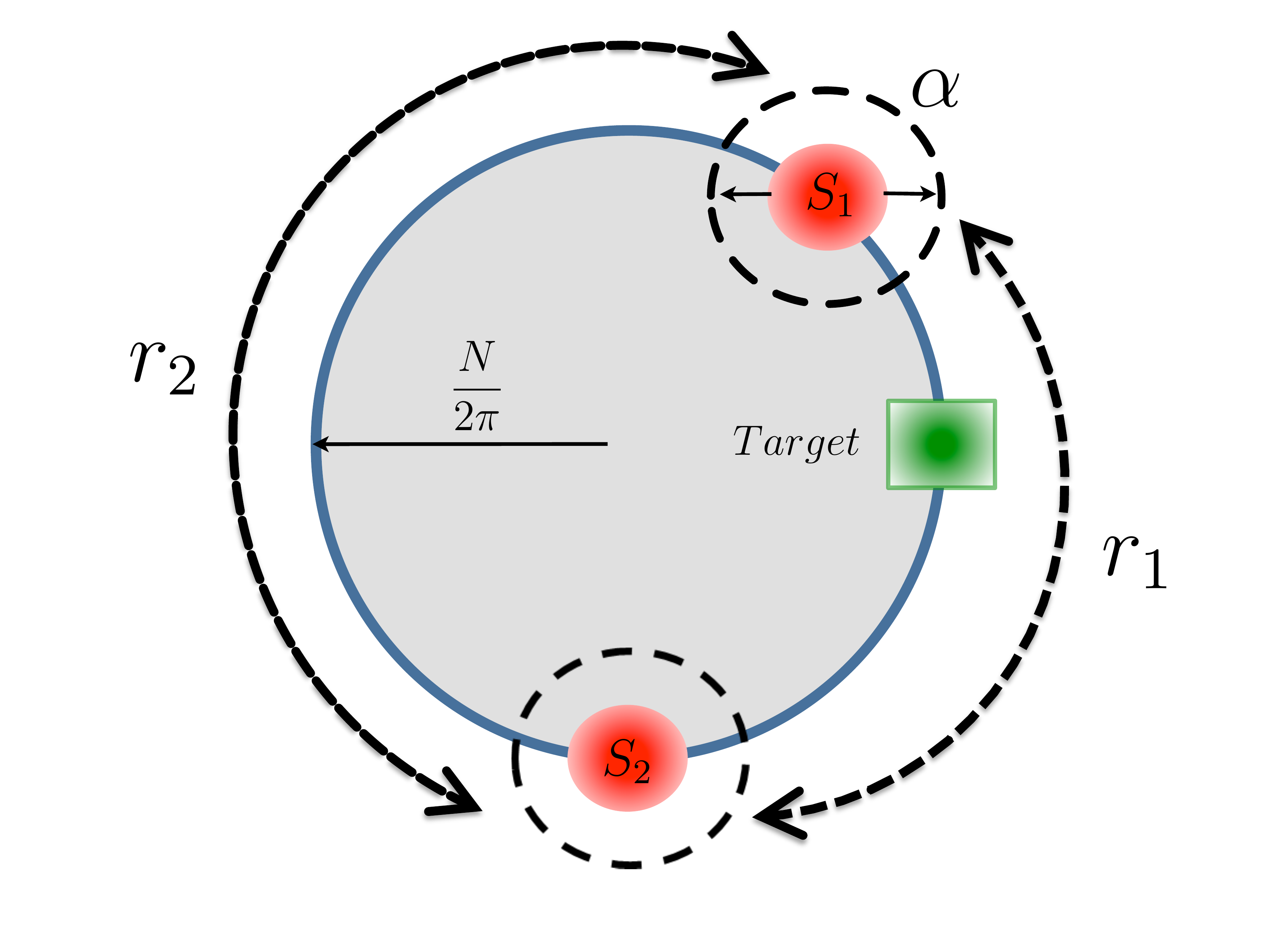}
\caption{The pictorial representation of our simulation model. $S_1$ and $S_2$ are the positions of the two searchers (red circles), the green square is the target and the black dotted circles are the repulsive boundary set by the value of $\alpha$. $r_1$ and $r_2$ are the distances between the two searchers, and $\frac{N}{2\pi}$ is the effective radius of periodic system.
\label{fig:model}}
\end{figure}

We study a discrete system consisting of two interacting Brownian searchers (random walkers) that move along a one dimensional periodic lattice with $N$ sites (Fig.~\ref{fig:model}). Initialization of both searchers and the target are selected from a uniform random distribution, such that the domain of the distribution corresponds to the lattice size, $N$. The dynamics of this model are such that the bare diffusion constant for both searchers when they are not interacting is, 

\begin{equation}
D=\frac{a^{2}}{T}=1, 
\end{equation}

where the lattice spacing is $a=1$.\\ 

Repulsive interactions are considered only between the two mobile searchers. Specifically we use an inverse power of the distance between the two searchers similar in form to an electrostatic potential between two like charges. The form of the potential in general is $V=\alpha/r^\gamma$ and for our simulations $\gamma=1$. The range of the potential is set by the parameter $\alpha$ and in our simulations is given in terms of the lattice spacing $a$, which can range from $0$ to $2N$. The distance between the two searchers is given as,

\begin{equation}
r_{1} =\big \vert S_1 -S_2\big \vert, \label{eq:eq2}
\end{equation}

\begin{equation}
r_{2} = N - r_{1}.\label{eq:eq3}
\end{equation}

Here the $|\hspace*{3pt}|$ represents the absolute value and the subscripts refer to the relative distance between the searchers on either side of the periodic boundaries (Fig.~\ref{fig:model}). The distance to the target relative to either of the searchers is only used to end the simulation when one of the searchers``finds" the target. Once the target is found we record the time (i.e the number of time steps) it took for that specific realization of the simulation to end. Searchers' positional updates were found by evaluating the energy difference between the current system configuration and  
a randomly chosen proposed new configuration of the system ($S_i(t+1) = S_i(t) \pm a$). In this way the system has no memory beyond the previous time step. The energies of either configuration is given as,

\begin{equation}
E_{i}\:=\:\frac{\alpha}{r_{1,i}}+\frac{\alpha}{r_{2,i}}\label{eq:eq4}
\end{equation}

\begin{equation}
E_{f}\:=\:\frac{\alpha}{r_{1,f}}+\frac{\alpha}{r_{2,f}},\label{eq:eq5}
\end{equation}
where $i$ refer to the previous configuration before the searcher's positional update and $f$ refers to the new proposed configuration. Once these energies are calculated we then calculate the probability for the change in the configuration of the system using a Boltzmann distribution.

\begin{equation}
P\:=\exp(-(E_{f}-E_{i}))\label{eq:eq6}
\end{equation}

Employing the METROPOLIS Monte Carlo (MMC) method the value of $P$ is then compared to a random number $s$, which is drawn from a uniform distribution between $0$ and $1$. If $s<P$, then the proposed new configuration of the system is accepted, which corresponds to a lower energy of the system.  This procedure is carried out until the target is found. In light of the fact that will be measuring the MFPT of our system, the choice of MMC over the usage of kinetic Monte Carlo (kMC) is worth a brief discussion. Although kMC explicitly measures the dynamical transition rates of our simulation, which are related to the real passage times of the system, in this case, time as calculated via kMC turns out to be equivalent to the
actual computer simulation time, measured in MC simulation iteration steps, for sufficiently long simulation times . All statistical quantities that are presented here were computed by averaging over many initial conditions of the system for a fixed lattice size.

\section*{2. Results}
A single particle diffusing on a closed d-dimensional space has been studied in great detail. Here we focus on a random walk of two particles on a $1d$ closed manifold, or a ring where the two particles interact via a repulsive interaction. 

\subsection*{Results I - Distribution of first passage times}
\begin{figure}[ht]
\centerline{
\includegraphics[trim=10 50 50 150, clip=true, scale=0.25]{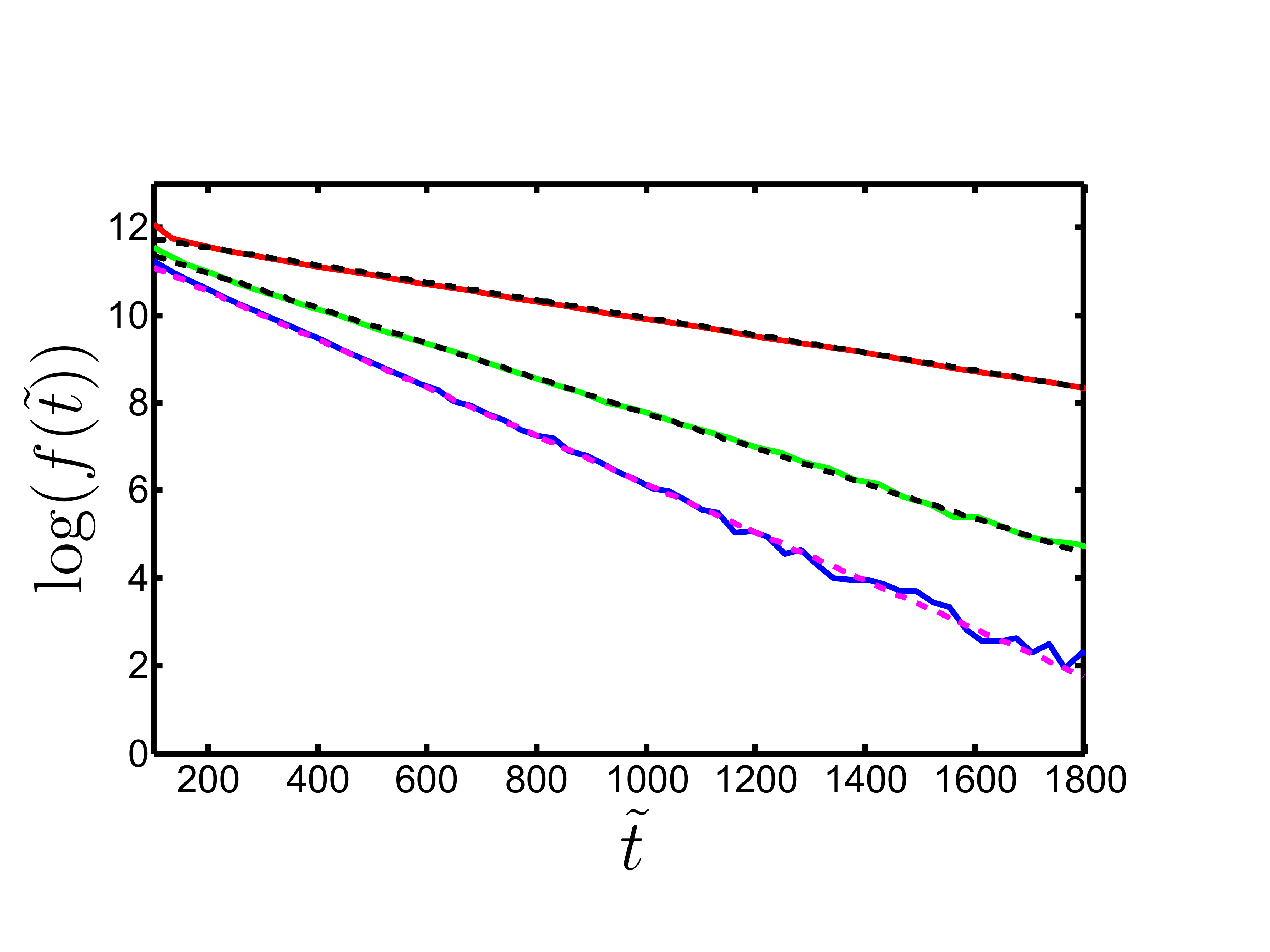}
}
\caption{ Semi-log plot of the frequency of search times for the single searcher (red solid line), two non-interacting searchers (green solid line) and 
two interacting searchers (blue solid line). The slope of the fitted dashed lines is the MFPT for each system and for the single searcher is $ \tau ^{-1}_{1,sim} =0.002$, for the two non-interacting searchers $\tau^{-1}_{2,sim} =0.004$ and for the two-interacting searchers is $\tau^{-1}_{2i,sim} =0.0005$.}
\label{fig:hist}
\end{figure}

To compare our results with previous studies for non-interacting Brownian searchers on a $1d$ ring, we calculate numerically the distribution of first passage times for a single searcher, two non-interacting searchers and two interacting searchers to reach the target over all random realizations of the target and searcher starting positions. Intuitively, one should expect that the time for repulsive searchers, on average, requires less time than the non-interacting cases. Since the repulsive interaction encourages the two searchers to avoid covering the same locations on the lattice, when their repulsive energies are of order $N/2$.  In Fig.~\ref{fig:hist} we plot the logarithm of the frequency of target encounters verses the time that was taken to reach the target (first passage times) for all three cases and find that indeed the repulsive random walkers finds the target faster (pink dashed line).  The first two cases, for the single searcher and two non-interacting searchers, we should expect that the mean first passage time to the target's position, averaged over all initial starting positions of the target, to be proportional to the square of the system size for large lattice sizes. More precisely we should expect for finite lattices that the mean number of steps taken before the target is found is~\cite{Montroll}

\[ \langle n \rangle =N(N+1)/6.\]

In our simulations, for the case when the searchers are non-interacting, we have that the number of searcher steps taken equals the number of time steps in our simulation before the target is found, hence $\langle \tau \rangle = \langle n\rangle$. In Fig.~\ref{fig:hist} we find an excellent agreement with our numerical simulation for both the single random walker and two non-interacting walkers, which yield a theoretical MFPT for a system size of N=50 as $\langle \tau \rangle^{-1}_1 =0.002$ and  $\langle \tau \rangle^{-1}_2 =0.004$ (black dashed lines). 

\begin{figure}[h]
\centerline{\includegraphics[trim=10 100 70 200, clip=true, scale=0.27]{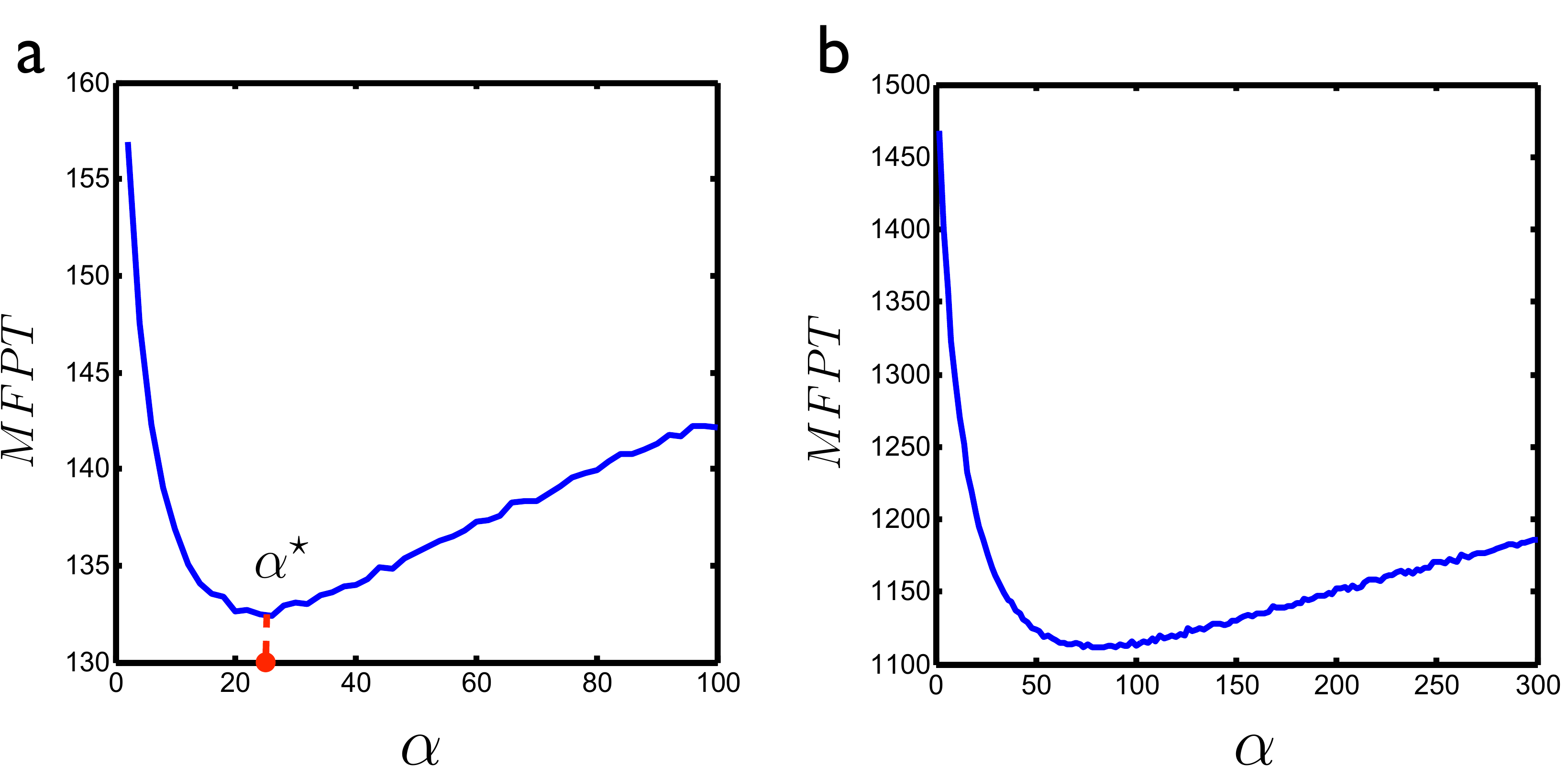}}
\caption{\textbf{a}. The MFPT over the range of repulsive strengths $\alpha = [0,2N]$ for a lattice size of $N=50$. The red marker indicates approximately where the minimum MFPT occurs at a point $\alpha^{\star}$. \textbf{b}. The MPFT verses repulsive strength for a system size of $N=150$. Both figure clearly show a minimum MFPT, but the location of $\alpha^{\star}$ changes as the system size is changes.}
\label{fig:two-graph}
\end{figure}

We find in general that the addition of a repulsive interaction lowers the mean first passage time in comparison to the non-interacting cases. We do expect that as we approach the limit where $\alpha$ goes to zero that we should recover the mean first passage time for the two non-interacting searchers. However, in the limit that $2\alpha/N$ becomes comparable to the thermal noise in the system ($k_b T=1$), we should expect that either searcher exhibits slower effective diffusive motion ($\text{number of steps}^2/\text{total time}<1$). This implies that the mean first passage time to the target should increase, thus we should expect to find an optimal repulsive strength where the searchers discover the target in the least amount of time.

\begin{figure*}[!tbh]
\centerline{\includegraphics[trim=10 70 0 25, clip=true, scale=0.45]{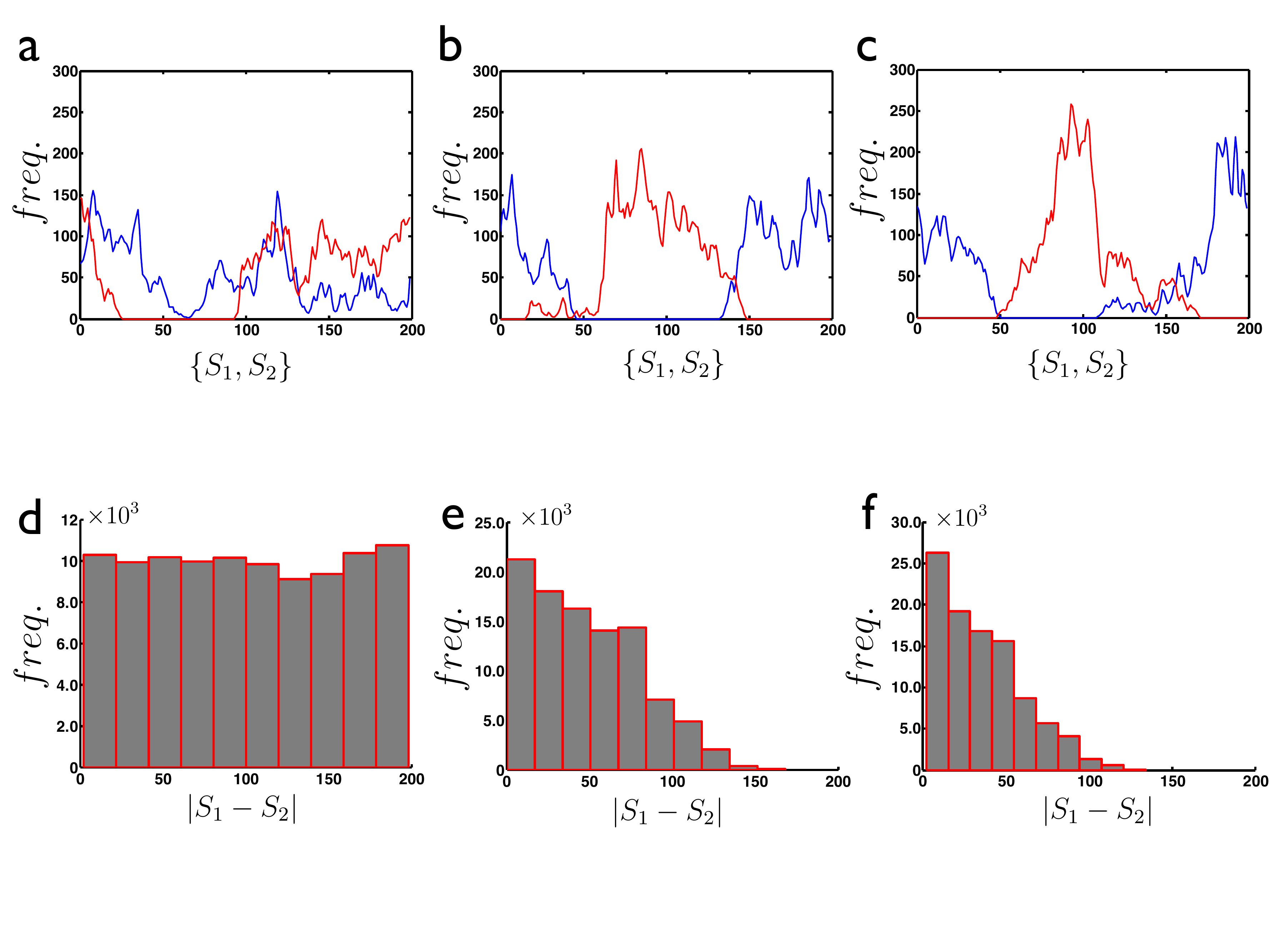}}
\caption{\textbf{a}. The frequency of lattice position occupation for both searchers (red-$S_1$) and (blue-$S_2$) for $\alpha=0$ displays a high value of overlap. \textbf{b}. At the optimal value of the repulsion strength $\alpha^{\star}\approx N/2$ we find that the overlap between the two searchers is minimal while maintaining a good search coverage over the lattice. \textbf{c}. Past the optimal repulsive strength we find that while the overlap is minimal, the search coverage of the lattice is poor because the searchers are more localized. \textbf{d-f}. The frequency of particle separation as a function of the absolute value of the relative coordinate $\vert S_1 -S_2\vert$. For the non -interaction case all searcher separations are equally likely, consistent with the overlap shown in Fig. \textbf{a}. In contrast, at the optimal and large repulsion values (as in Fig. \textbf{b-c})  we find that the most likely separation is when $S_1\approx S_2$, which is consistent with the optimal repulsion value of $\alpha \approx N/2$.
All data shown was for a lattice size of $N=200$. For \textbf{a-c} simulations were run for $10^5$ time steps. For \textbf{d-f} simulations were run for $10^6$ time steps.}
\label{fig:freq-lat-pos-rep}
\end{figure*}

\subsection*{Results II - Repulsive Strength and Lattice Size}

To test the assertion that there exists an optimal repulsive strength, we performed simulations over the entire range of repulsive strengths such that the repulsive energy goes from $\alpha=[0, 2N]$. Fig.~\ref{fig:two-graph} shows ensemble averaged data for the mean first passage time as a function of the repulsive strength for two different lattice sizes. In each case we find that there is indeed a minimum mean first passage time that occurs at an optimal repulsive strength near $\alpha\approx N/2$. This result is intuitively correct since this is the repulsive range at which the two searchers begin to interact more strongly. Increasing the repulsion leads to an increase in the mean first passage time and a decrease in the effective diffusion rate for either of the two searchers. 

\begin{figure}[!tbh]
\centerline{
\includegraphics[trim=100 0 150 0, clip=true,  scale=0.3]{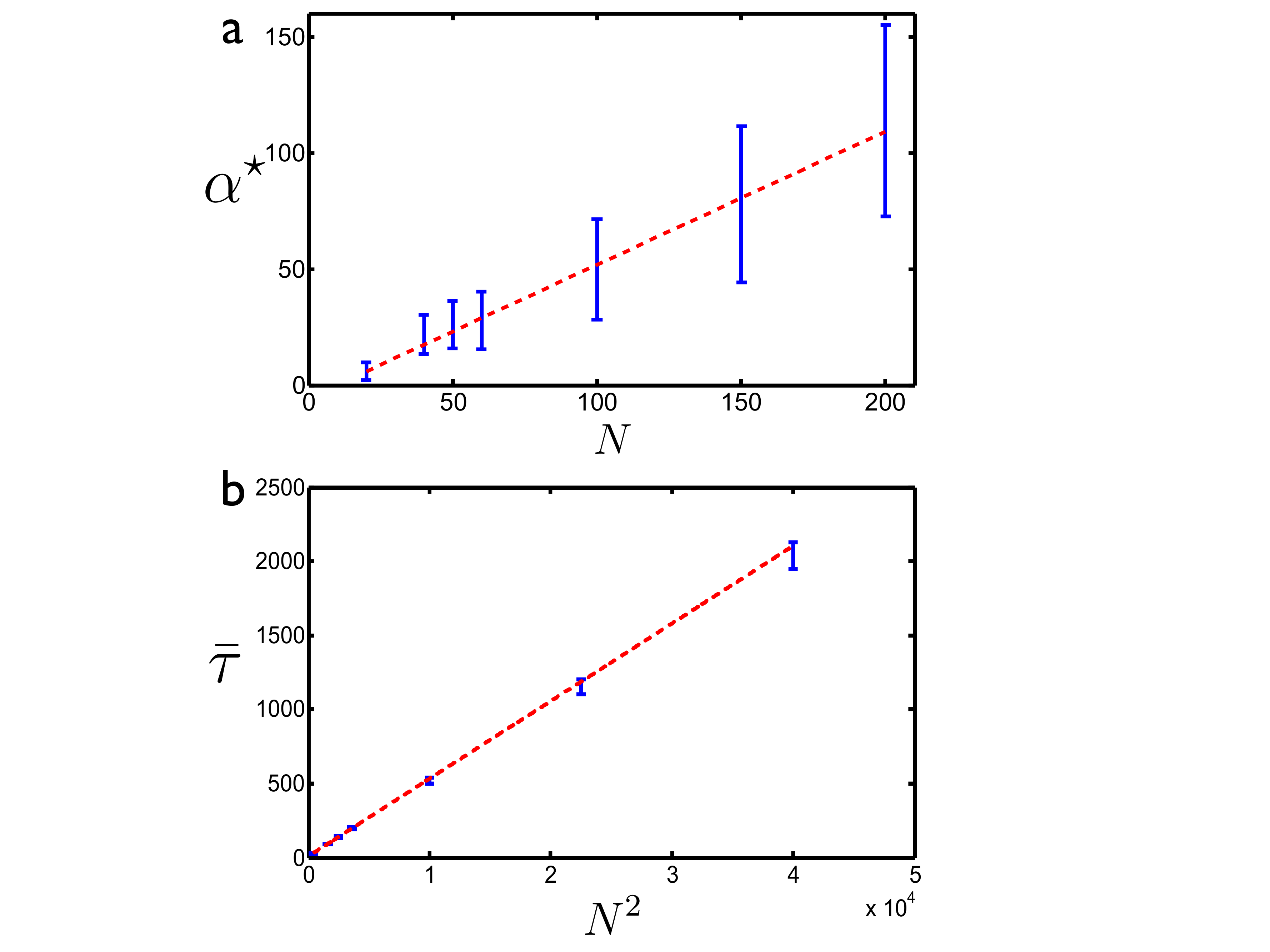}
}
\caption{\textbf{a}. Graph of lattice size to repulsive strength. Blue bars represent the error and red line is the 
best fit line $\alpha^{\star}=0.5755 \pm 0.05$ weighted by the error for each data point shown. Error bars were calculated using data points
that were within 2\% of the minimum (Fig.~\ref{fig:two-graph}) for each lattice size. \textbf{b}. Graph of $N^2$ and $\bar{\tau}$. Blue bars represent the error for each data point and the red line is the best fit line using to our scaling ansatz, which gives $\bar{\tau}=0.0507*N^{2}$} \label{fig:interp}
\end{figure}

At low values of the repulsion strength we find that the trajectories for each searcher overlaps frequently (Fig.~\ref{fig:freq-lat-pos-rep}a). As the repulsion strength is increased we find that the amount of overlap between the two searchers' trajectories becomes significantly less  (Fig.~\ref{fig:freq-lat-pos-rep}b) near the optimum shown in Fig.~\ref{fig:two-graph}. Increasing the repulsive interaction past the optimum we find that any overlap of the two trajectories is suppressed(Fig.~\ref{fig:freq-lat-pos-rep}c). In this regime, the energy associated with such a large value of $\alpha$ makes any over lap unfavorable during the simulation. To understand this, we can examine the allowed updates modes to the system (there are four in total) as the repulsion is increased. We should expect that allowed movements of the two searchers becomes restricted to particular set of  modes allowed by the partition function. There are two translational modes, where the change in the distance between the two searchers is unchanged $\Delta r_i = 0$, hence the energy of the system is unchanged. In contrast there exists two modes where the energy of the system changes, such that the searchers move apart or move closer with respect to each other, $\Delta r_i \neq 0$. For values of repulsion near or above the optimum, modes where $\Delta r_i \neq 0$ are unfavorable.\\

In Figs.~\ref{fig:freq-lat-pos-rep}d-f we show the frequency of searcher separations using the relative coordinate, $r_1$ modulo $N$, for a given repulsive strength ($\alpha = [0, N/2, N] \text{ respectively}$). For the case with no repulsion we find that all possible particle separations are possible, hence the flat distribution in Fig.~\ref{fig:freq-lat-pos-rep}d. As we turn up the repulsive interaction ($\alpha =N/2$ (Fig.~\ref{fig:freq-lat-pos-rep}e) and $\alpha=N$ (Fig.~\ref{fig:freq-lat-pos-rep}f)) we find that the distribution of allowed particle separations is limited to only a set that is maximal at $r_1=0$. This observation suggest that for repulsive strengths that are near or above the optimal that the searchers are more likely to be found at diametrically opposed locations on the lattice with very little over lap of their search area, which is consistent with Figs.~\ref{fig:freq-lat-pos-rep}a-b.

\subsection*{Results III - Average Encounter Time and Lattice Size}
To shed light on the reason for the optimal value of the repulsive interaction, one can use dimensional analysis to determine the location optimal value of $\alpha$ as a function of the system size. The are two important length scales in our system, one of which is the lattice size $N$ and the other is determined by the diffusion constant \textit{D} (length\textsuperscript{2}/time). We posit that the critical (optimal) value of the repulsive energy is a function of both of these parameters such that, $\alpha^{\star}=f(N,D)$. Since the repulsive energy is dimensionless (Eqns. 4 and 5), $\alpha$ is defined to have units of length. The bare diffusion of the two searchers is defined in terms of the underlying length and time scales of our simulation and is unity. Therefore we expect the form of the scaling function to be linear in the lattice size, 
\[
\alpha^{\star}=\: bN,
\]
where $b$ is a unit-less constant. 

In Fig~\ref{fig:interp} we plot the location of the optimal value of $\alpha$ verses the system size $N$. Fitting the data using the linear fitting function suggested by our dimensional scaling analysis, we find an excellent agreement between the data and the scaling function (red dashed line) proposed by our dimensional analysis. The fitting free parameter $b$ is found to be about 0.6 which is very close to what we expect intuitively since at the value of $\alpha=N/2$ is where the two searchers begin to interact at an energy scale where the probability for accepting/rejecting a proposed move is of order $\sim e^{-1}$. This optimal repulsive strength indicates that the interaction between the searchers are such that both do not diffuse over the same space (Fig.~\ref{fig:freq-lat-pos-rep}b) as doing so would be energetically unfavorable.

In addition to the existence of an optimal repulsive interaction, which lowers the average mean first passage time to the target, we also expect there will be an average (minimal) encounter time to the target, $\bar{\tau}$. This characteristic time is related to the intrinsic length and time scales that define our system and should also be related to other relevant quantities, like the bare diffusion constant, $D$. Using dimensional analysis, we can write down a scaling ansatz similar to the one for the optimal repulsive strength $\alpha^{\star}$. Since we are interested in a quantity that has units of \textit{time}, then the only two parameters that the mean encounter time can depend on is the lattice size and the bare diffusion rate, $T\:=f(N,D)$. Given that the diffusion constant has units of $[Length^{2}*Time^{-1}]$ we expect that the dependence of $\bar{\tau}$ to scale as the square of the system size. Therefore we expect that,
\[
\bar{\tau}=a*\frac{N^{2}}{D}.
\]
In Fig.~\ref{fig:interp}b we find the average mean first encounter time to the target as a function of the lattice size squared (blue points). Fitting this data with the scaling function derived above we found an excellent agreement between the data and the best fit line (red dashed line).  

\section*{3. Discussion}

In summary, we have shown that for two Brownian searchers foraging for a single target that the mean first passage time (MFPT) to the target is minimized when there is a mutually repulsive interaction between them. We have also shown that the optimal repulsive range between the two searchers is roughly the maximal separation between searchers. These results suggest that the mutually repulsive interaction introduces a cooperative effect that allows the two searchers to optimize both their search time as well as the search area. Minimizing the search time by each searcher translates to real systems as a minimization of energy consumed by each searcher when foraging for food. This scheme is certainly realized in biological systems that utilize chemotactic signaling such as Dictyostelium (Dicty). When food supplies are low single cell Dicty begins to produce chemo-repellent in places that it has searched for food~\cite{Keating,Konijn,Pan}. This has the effect of informing other single Dicty cells in the population to avoid these areas as way to conserve energy. This strategy allows the entire population to search for resources collectively as opposed to when food supplies are abundant, where each cell is free to search randomly. From this point of view it is easy to see why having an antagonist interaction among foragers is evolutionarily advantageous, by minimizing energy consumption Dicty can ensure that it's fitness in a nutrient poor environment will remain high~\cite{Gelbart}. This result opens an interesting evolutionary question in that not only do organisms have to utilize some intercommunication signal to optimize their resource gathering but that interaction must all be optimally tuned in order for collective foraging to always promote a high fitness value in a wide range of environmental conditions. 

\section*{Acknowledgments}
The authors would like to thank both Anatoly Kolomeisky and KC Huang for their insightful comments and discussions.  
This work was partially supported by National Science Foundation grant EF-1038697 and a James S. McDonnell Foundation Award.
Undergraduate support was funded by the Undergraduate Research and Mentoring (URM) Program Sponsored by the National Science Fundation NSF Grant DBI-1040962.

\newpage
\bibliography{ref}

\end{document}